\documentstyle[preprint,aps,epsf]{revtex}

\global\let\epsfloaded=Y

\begin{document}
\draft
\title{Strong and Electroweak Interactions, 
and Their Unification with Noncommutative
Space-Time}
\author{Xiao-Gang He}
\address{Institute of Theoretical Physics, Academia Sinica, Beijing 100080\\
 and\\ 
Department of Physics, National Taiwan University,
Taipei 10764}
\maketitle

\begin{abstract}
Quantum field theories based on noncommutative space-time (NCQFT) 
have been extensively studied
recently. However no NCQFT model, which can
uniquely describe the strong and electroweak interactions, 
has been constructed. This 
prevents consistent and systematic study of 
noncommutative space-time. In this work we construct a NCQFT model based
on the trinification gauge group $SU(3)_C\times SU(3)_L\times SU(3)_R$.  
A unique feature of this model, that 
all matter fields (fermions and Higgses) are assigned to
(anti-)fundamental representations of the factor $SU(3)$ groups,
allows us to construct a NCQFT model for strong and 
electroweak interactions and their unification without ambiguities. 
This model provides an example which allows
consistent and systematic study of
noncommutative space-time phenomenology. 
We also comment on some related issues regarding extensions to $E_6$ 
and $U(3)_C\times U(3)_L\times U(3)_R$ models.
\end{abstract}


\pagestyle{empty} 
%
%
\vfill

\vfill

%
%
\vfill

%
%

%
%
\pagestyle{plain} Noncommutative quantum field theory (NCQFT), based on
modification of the space-time commutation relations, provides an
alternative to the ordinary quantum field theory. A simple way to modify the
space-time properties is to change the usual space-time coordinate $x$ to
noncommutative coordinate $\hat X$ such that\cite{1}

\begin{eqnarray}
[ \hat{X}^{\mu },\hat{X}^{\nu }]=i\theta ^{\mu \nu },
\end{eqnarray}
where $\theta ^{\mu \nu }$ is a real anti-symmetric matrix. We will consider
the case where $\theta ^{\mu \nu }$ is a constant and commutes with $\hat{X}%
^{\mu }$.

NCQFT based on the above commutation relation can be easily studied using
the Weyl-Moyal correspondence replacing the product of two fields $A(\hat X)$
and $B(\hat X)$ with noncommutative coordinates by product of the same
fields but ordinary coordinate $x$ through the star ``$*$'' product,

\begin{eqnarray}
A(\hat X) B(\hat X) \to A(x)*B(x) = Exp[i {\frac{1}{2}} \theta^{\mu\nu}
\partial_{x,\mu} \partial_{y,\nu}] A(x) B(y)|_{x=y}.
\end{eqnarray}

Properties related to NCQFT have been studied extensively 
recently\cite{2,3,3a,4,5,6,7,8,9,10,11,12}.
NCQFT for a pure 
$U(1)$ group is easy to study. Related phenomenology have been
studied recently\cite{2}. But it is more complicated for non-abelian groups.
Due to the ``$*$'' product nature, there are fundamental differences between
ordinary and noncommutative gauge theories 
and cause many difficulties to 
construct a unique and consistent model for strong and electroweak 
interactions based on $SU(3)_C\times SU(2)_L\times U(1)_Y$ gauge 
group\cite{3,3a,4,5,6,7,8,9,10,11,12}.  
One of the main problems is that $SU(N)$ group can not be simply gauged 
with ``$*$'' product as will be explain in the following.
Another problem is that, naively, 
the charges of any $U(1)$ gauge group with ``$*$'' product 
are quantized to only three 
possible values, 1, 0, -1 which cannot accommodate all the hypercharges 
for matter fields in the SM.

In this work we construct a NCQFT model based
on the trinification gauge group $SU(3)_C\times SU(3)_L\times SU(3)_R$.  
We show that NCQFT model for strong and electroweak
interactions and their unification can be consistently constructed.
This model therefore provides an example which allows a consistent and 
systematic study of noncommutative space-time phenomenology. 

With noncommutative space-time there are modifications to the fields compared 
with the ordinary ones. We
indicate the fields in NCQFT with a hat and the ordinary ones without hat. 
The definition of gauge
transformation $\hat \alpha$ of a gauge field $\hat A_\mu$ for a $SU(N)$ 
is similar to the ordinary one 
but with usual product replaced by the ``$*$''
product. For example

\begin{eqnarray}
\delta_\alpha \hat \phi = i \hat \alpha * \hat \phi,
\end{eqnarray}
where $\hat \phi$ is a fundamental representation of $SU(N)$. We use the
notation $\hat A_\mu = \hat A^a_\mu T^a$, $\hat \alpha = \alpha^a T^a$
with $T^a$ being the $SU(N)$ generator normalized as 
$Tr(T^aT^b) = \delta^{ab}/2$.

Due to the noncommutativity of the space-time, 
two consecutive local transformations $\hat{\alpha}$ and $\hat{\beta}$
of the type in the above,

\begin{eqnarray}
(\delta_\alpha \delta_\beta - \delta_\beta\delta_\alpha) = (\hat \alpha*\hat 
\beta - \hat \beta * \hat \alpha),
\end{eqnarray}
cannot be reduced to the matrix commutator of the generators of the Lie
algebra due to the noncommutativity of the space-time. They have to be in
the enveloping algebra

\begin{eqnarray}
\hat \alpha = \alpha + \alpha^1_{ab} :T^aT^b: + ... +
\alpha^{n-1}_{a_1...a_n}:T^{a_1}...T^{a_n}: +...
\end{eqnarray}
where $:T^{a_1}...T^{a_n}:$ is totally symmetric under exchange of $a_i$.
This poses a difficulty in constructing non-abelian $SU(N)$ 
gauge theories\cite{3}.

Seiberg and Witten have shown\cite{4} that the fields defined in
noncommutative coordinate can be mapped on to the ordinary fields, the
Seiberg-Witten mapping. In Ref.\cite{5} it was shown that this mapping
actually can be applied to the ``$*$'' product with any gauge groups. It is
possible to study non-abelian gauge group theories. Using the above
enveloping algebra, one can obtain the noncommutative fields in terms of the
ordinary fields with corrections in powers of the noncommutative parameter, 
$\theta^{\mu\nu}$, order by order. To the first order in $\theta^{\mu\nu}$
noncommutative fields can be expressed as

\begin{eqnarray}
&&\hat \alpha = \alpha +{\frac{1}{4}} \theta^{\mu\nu}\{\partial_\mu \alpha, 
A_\nu\}
+ c \theta^{\mu\nu}[\partial_\mu \alpha, A_\nu],  \nonumber \\
&&A_\mu = -{\frac{1}{4}} \theta^{\alpha\beta}\{A_\alpha, \partial_\beta
A_\mu + F_{\beta\mu}\} + c \theta^{\alpha\beta}([A_\alpha,\partial_\mu
A_\beta] +i[A_\alpha A_\beta, A_\mu]),  \nonumber \\
&&\hat \phi = a\theta^{\mu\nu} F_{\mu\nu} \phi -{\frac{1}{2}}
\theta^{\mu\nu} A_\mu\partial_\nu \phi + i({\frac{1}{4}} + c)\theta^{\mu\nu}
A_\mu A_\nu\phi,  \label{ncf}
\end{eqnarray}
where $F_{\mu\nu} = \partial_\mu A_\nu - \partial_\nu A_\mu -ig_N[A_\mu,
A_\nu]$. The term proportional to $a$ can be absorbed into the redefinition
of the matter field $\phi$. The parameter $c$ can not be removed by
redefinition of the gauge field. It must be a purely imaginary number from
the requirement that the gauge field be self conjugate.

Using the above noncommutative fields, one can construct a gauge theory 
for $SU(N)$ group. The action $S$ of a $SU(N)$ NCQFT, to the leading order
in $\theta$, is
given by\cite{5,6},

\begin{eqnarray}
S &=&\int Ld^{4}x,  \nonumber \\
L &=&-{\frac{1}{2}}Tr(F_{\mu \nu }F^{\mu \nu })+{\frac{1}{4}}g_{N}\theta
^{\mu \nu }Tr(F_{\mu \nu }F_{\rho \sigma }F^{\rho \sigma }-4F_{\mu \rho
}F_{\nu \sigma }F^{\rho \sigma })  \nonumber \\
&+&\bar{\phi}(i\gamma ^{\mu }D_{\mu }-m)\phi -{\frac{1}{4}}\theta ^{\alpha
\beta }\bar{\phi}F_{\alpha \beta }(i\gamma ^{\mu }D_{\mu }-m)\phi -{\frac{1}{%
2}}\theta ^{\alpha \beta }\bar{\phi}\gamma ^{\mu }F_{\mu \alpha }iD_{\beta
}\phi ,  \label{ss}
\end{eqnarray}
where $D_\mu =\partial_{\mu }-ig_{N}T^{a}A_{\mu }^{a}$. We note that the
parameter $c$ does not appear in the Lagrangian. The Lagrangian is
uniquely determined to order $\theta $. We will therefore work with a simple
choice $c=0$ from now on. In the above, if $\phi $ is a chiral field, $m=0$.

To obtain a theory which can describe the strong and electroweak
interactions such as the Standard Model (SM), one also needs to solve 
the $U(1)$ charge quantization problem, namely only
three possible values, 1, 0, -1 for $U(1)$ charges, as mentioned earlier.
It has been shown that this difficulty can also be overcome with the use of 
the Seiberg-Witten mapping\cite{4} .

To solve the $U(1)$ charge quantization problem, one associates each charge $%
gq^{(n)}$ of the nth matter field a gauge field $\hat A_\mu^{(n)}$\cite{7}.
In the commutative limit, $\theta^{\mu\nu} \to 0$, $\hat A_\mu^{(n)}$
becomes the single gauge field $A_\mu$ of the ordinary commuting space-time $%
U(1)$ gauge theory. But at non-zero orders in $\theta^{\mu\nu}$, $\hat A%
^{(n)}_\mu$ receives corrections\cite{7}. 
In doing so, the kinetic energy of the gauge boson will, however, be
affected. Depending on how the kinetic energy is defined (weight over
different field strength of $\hat A_\mu^{(n)}$), the resulting kinetic
energy will be different even though the proper normalization to obtain the
correct kinetic energy in the commutative limit is imposed\cite{7}.

In the SM there are six different matter field multiplets for each
generation, i.e. $u_R$, $d_R$, $(u,\;d)_L$, $e_R$, $(\nu,\;e)_L$ and $%
(H^0,\;H^-)$, a priori one can choose a different $g_i$ for each of them.
After identifying three combinations with the usual $g_3$, $g_2$ and $g_1$
couplings for the SM gauge groups, there is still a freedom to choose
different gauge boson self interaction couplings at non-zero orders in $%
\theta^{\mu\nu}$. This leads to ambiguities in self interactions of gauge
bosons when non-zero order terms in $\theta^{\mu\nu}$ are included\cite{7}.
This problem needs to be resolved. A way to solve this problem is to have a
theory without the use of $U(1)$ factor group.

There are many groups without $U(1)$ factor 
group which contain the SM gauge group
and may be used to describe the strong and electroweak interactions. However
not all of them can be easily extended to a full NCQFT using the formulation
described above. For example one can easily
obtain unique gauge boson self interactions in $SU(5)$ theory\cite{8}. But
the matter fields require more than one representations 5 and 10 which causes
additional complications\cite{9} and will reintroduce the uniqueness problem 
for kinetic energy. One way of obtaining 
a consistent NCQFT is to  have a theory in which all matter and Higgs 
fields are in the same representation such that once Seiberg-Witten mapping is 
used to solve the problem of gauging a $SU(N)$ group, there is 
no problem 
with unique determination of kinetic energy. To
this end we propose to use the trinification gauge group\cite{13}, $%
SU(3)_C\times SU(3)_L\times SU(3)_R$ with a $Z_3$ symmetry.

This theory leads to unification of strong and electroweak interactions. An
important feature of this theory is that the matter and Higgs fields are
assigned to (anti-)fundamental representations of the factor $SU(3)$ groups
and therefore the formalism described earlier can be readily used.

In the trinification model, the gauge fields are in the adjoint
representation,

\begin{eqnarray}
24= A^C + A^L + A^R = (8,1,1) + (1,8,1) + (1,1,8),
\end{eqnarray}
which contains 24 gauge bosons. $A^C$ contains the color gluon bosons, a
linear combination of component fields of $A^L$ and $A^R$ forms the $U(1)_Y$
gauge boson, and $A^L$ contains the $SU(2)_L$ gauge bosons. The rest are
integrally charged heavy gauge bosons which do not mediate proton 
decays\cite{13}.

Each generation of fermions is assigned to a 27,

\begin{eqnarray}
\psi &=& \psi^{LR} + \psi^{RC} + \psi^{CL} = (1,3,\bar 3) + (\bar 3,1,3) +
(3,\bar 3,1),  \nonumber \\
\psi^{LR} &=& \left ( 
\begin{array}{lll}
E^0 & E^- & e^- \\ 
E^+ & \bar E^0 & \nu \\ 
e^+ & N_1 & N_2
\end{array}
\right ),\;\; \psi^{RC} = \left ( 
\begin{array}{lll}
\bar u_1 & \bar u_2 & \bar u_3 \\ 
\bar d_1 & \bar d_2 & \bar d_3 \\ 
\bar B_2 & \bar B_2 & \bar B_2
\end{array}
\right ),\;\; \psi^{CL} = \left ( 
\begin{array}{lll}
u_1 & d_1 & B_1 \\ 
u_2 & d_2 & B_2 \\ 
u_3 & d_3 & B_3
\end{array}
\right ).
\end{eqnarray}
In the above we have written the fermions in left-handed chiral fields. The 
$B$ field is a heavy particle.

The Higgs fields which break the trinification to the SM gauge group are
also assigned to 27 representations. In order to have correct mass patterns,
at least two 27 Higgs representations are needed\cite{13}. We indicate them
by

\begin{eqnarray}
\phi_i = \phi_i^{LR} + \phi_i^{RC} + \phi_i^{CL} = (1,3,\bar 3)_i + (\bar 3,
1,3)_i + (3, \bar 3,1)_i.
\end{eqnarray}

The $Z_3$ symmetry operates in the following way. If $(C,L,R)$ is a
representation under the $SU(3)_C\times SU(3)_L\times SU(3)_R$, the effect
of $Z_3$ is to symmetrize it to

\begin{eqnarray}
Z_3(C,L,R) = (C,L,R) + (R,C,L) + (L,R,C).
\end{eqnarray}
Requirement of the Lagrangian to be invariant under $Z_3$ relates the gauge
couplings $g^{C,L,R}$ of the gauge groups to be equal, $g^C=g^L=g^R=g^U$, at
a scale which is the unification scale of the model.

The vacuum expectation values of the Higgs scalars break the symmetry to 
$SU(3)_C\times SU(2)_L \times U(1)_Y$ have the following form

\begin{eqnarray}
<\phi^{LR}_1> = \left ( 
\begin{array}{lll}
\hat 0 & 0 & 0 \\ 
0 & \hat 0 & 0 \\ 
0 & 0 & v_1
\end{array}
\right ),\;\; <\phi^{LR}_2> = \left ( 
\begin{array}{lll}
\hat 0 & 0 & 0 \\ 
0 & \hat 0 & \hat 0 \\ 
0 & v_2 & \hat 0
\end{array}
\right ).
\end{eqnarray}
Non-zero values of $v_{1,2}$ break the symmetry to the SM group. The scale
of $v_{1,2}$ are at the unification scale. At
this stage 12 of the gauge bosons, and $B$, $E_i$ and $N_i$ particles
receive masses. They are therefore very heavy. The entries indicated by 
$\hat 0$ can develop VEV's of order $m_W$. These VEV's break the SM group to 
$SU(3)_C\times U(1)_{em}$ and provide masses for the ordinary quarks and
leptons.

In this model $\sin^2\theta_W = 3/8$ at the unification scale. Using the
present electroweak precision test data for $\sin^2\theta_W$, $\alpha_s$ and
$\alpha_{em}$ at the $Z$ mass pole, the unification scale is determined to be
$10^{14}$ GeV\cite{will}. A unification scale as low as $10^{14}$ GeV in
a $SU(5)$ theory, for example, would induce rapid proton
decays and be ruled out. However, as have been mentioned previously that 
in the trinification
theory, gauge bosons do not mediate proton decays. Therefore the theory
would not have problem with proton decays. Mediation of heavy Higgs particles
can produce proton decays\cite{13}. However in this case, there are
many free parameters in the Yukawa and the Higgs potential couplings to make
the theory to be consistent with data\cite{13}.  

It is clear that the trinification model can provides an easy framework for
building a phenomenologically acceptable and 
consistent NCQFT model for strong and electroweak interactions. 
To the first order in $\theta^{\mu\nu}$, the
noncommutative gauge fields are the same form for the gauge fields as in eq.
(\ref{ncf}). The fermion and Higgs fields are in the same representation and
are all (anti-)fundamental representations $\phi$ of the type $(3,\bar 3)$
under the subgroups $SU(3)\times SU(3)$. The noncommutative fields expressed
in the ordinary fields are given by

\begin{eqnarray}
\hat \phi = \phi - {\frac{1}{2}} \theta^{\mu\nu} ( A_\mu \partial_\nu \phi - 
{\frac{i}{2}} A_\mu A_\nu \phi + \partial_\nu \phi A^{\prime}_\mu +{\frac{i}{%
2}}\phi A^{\prime}_\nu A^{\prime}_\mu),  \label{nncf}
\end{eqnarray}
where $A_\mu$ and $A^{\prime}_\mu$ are the gauge fields of the first $SU(3)$
and the second $SU(3)$ gauge groups, respectively.

The first order in $\theta^{\mu\nu}$ corrections to the Lagrangian $\tilde L$
for gauge and fermion kinetic energy terms are given by

\begin{eqnarray}
\tilde L &=& [{\frac{1}{4}}g^C \theta^{\mu\nu}Tr(F^C_{\mu\nu} F^C_{\alpha
\beta} F^{C\alpha\beta} -4F^C_{\alpha \mu} F^C_{\beta\nu} F^{C\alpha\beta})
+ (C\to L) + (C\to R)]  \nonumber \\
&-&[{\frac{i}{4}} \theta^{\mu\nu}Tr(\bar \psi^{LR} F^L_{\mu\nu}\gamma^\alpha
D_\alpha \psi^{LR} +F^R_{\mu\nu}\bar \psi^{LR}\gamma^\alpha D_\alpha
\psi^{LR}  \nonumber \\
&+&2\bar \psi^{LR} F^L_{\alpha\mu}\gamma^\alpha D_\nu \psi^{LR}
+2F^R_{\alpha\mu}\bar \psi^{LR}\gamma^\alpha D_\nu \psi^{LR}) + (LR \to LC)
+ (LR \to CR)],  \label{int}
\end{eqnarray}
where $D_\mu \psi^{LR} = \partial_\mu \psi^{LR} - ig_L A^L_\mu \psi^{LR} + i
g_R\phi^{LR} A^R_\mu$.

The above Lagrangian uniquely determine interactions due to noncommutative
space-time correction to the first order in $\theta ^{\mu \nu }$ without the
problems pointed out earlier. 
We emphasis that although the resulting theory at low energies appears
to have $U(1)$ factor group(s), the corresponding gauge self-interactions
are fixed because of the choice of trinification group which dictates
how gauge bosons interact.
From the above Lagrangian one can easily study
new interactions due to noncommutative space-time and test the model by
experimental data. For illustration, we present the neutral gauge boson self
interactions and its interactions with the SM fermions. Expanding the above
Lagrangian we obtain

\begin{eqnarray}
L_{int} &=&{\frac{1}{4}}g^{C}\theta ^{\mu \nu }Tr(G_{\mu \nu }G_{\alpha
\beta }G^{\alpha \beta }-4G_{\alpha \mu }G_{\beta \nu }G^{\alpha \beta }) 
\nonumber \\
&+&{\frac{1}{16}}\theta ^{\mu \nu }g_{Y}[c_{W}({\frac{7}{15}}
c_{W}^{2}+s_{W}^{2})(F_{\mu \nu }F_{\alpha \beta }F^{\alpha \beta
}-4F_{\alpha \mu }F_{\beta \nu }F^{\alpha \beta })  \nonumber \\
&-&s_{W}({\frac{7}{15}}s_{W}^{2}+c_{W}^{2})(Z_{\mu \nu }Z_{\alpha \beta
}Z^{\alpha \beta }-4Z_{\alpha \mu }Z_{\beta \nu }Z^{\alpha \beta }) 
\nonumber \\
&+&c_{W}(c_{W}^{2}-{\frac{23}{15}}s_{W}^{2})(F_{\mu \nu }Z_{\alpha \beta
}Z^{\alpha \beta }+2Z_{\mu \nu }Z_{\alpha \beta }F^{\alpha \beta
}-4(Z_{\alpha \mu }Z_{\beta \nu }F^{\alpha \beta }+2F_{\alpha \mu }Z_{\beta
\nu }Z^{\alpha \beta }))  \nonumber \\
&-&s_{W}(s_{W}^{2}-{\frac{23}{15}}c_{W}^{2})(Z_{\mu \nu }F_{\alpha \beta
}F^{\alpha \beta }+2F_{\mu \nu }Z_{\alpha \beta }F^{\alpha \beta
}-4(F_{\alpha \mu }F_{\beta \nu }Z^{\alpha \beta }+2Z_{\alpha \mu }F_{\beta
\nu }F^{\alpha \beta }))]
\end{eqnarray}
where $c_{w}=\cos \theta _{W}$, $s_{W}=\sin \theta _{W}$. $G_{\mu \nu }$, $%
F_{\mu \nu }$, $Z_{\mu \nu }=\partial _{\mu }Z_{\nu }- \partial_{\nu }Z_{\mu
}$ are the field strengths for the gluon, photon and Z particles,
respectively. Note that the above interactions are obtained at the
unification scale where $g_{Y}=\sqrt{3/5}g^{U}$ and $\sin ^{2}\theta
_{W}=3/8 $, and $g^{C}=g^{L}=g^{R}=g^{U}$.

From the above we see that the triple neutral gauge boson interactions are
uniquely determined unlike the case with SM gauge group studied in Ref. \cite
{7}. These interactions are also different from those predicted by $SU(5)$
model\cite{8}. This can be used to test the model\cite{10}.

The fermion-gauge boson interactions can readily be obtained by expanding
eq. (\ref{int}). The Yukawa coupling terms and Higgs potential terms can
also be obtained using results in eqs. (\ref{ncf}) and (\ref{nncf}).

We have constructed a NCQFT unification model of strong and electroweak
interactions based on $SU(3)_C\times SU(3)_L\times SU(3)_R\times Z_3$ group.
In this model all interactions are determined. New gauge boson self- and
fermion-gauge interactions are predicted. If the noncommutative scale turns
out to be low, the model can be tested experimentally. This model
provides an example which can consistently describe the strong and
electroweak interactions and their unification, and allows a systematic
investigation of the hypothesis of noncommutative space-time. 

Before closing we would like to make 
two comments on some possible extensions of the
model discussed here. One of them concerns about $E_6$ extension of the
model.
The $SU(3)_C\times SU(3)_L \times SU(3)_R$ group can be embedded into
$E_6$ group. One therefore can try to construct a NCQFT
based on $E_6$. 
With this group the gauge bosons are in the $78$ adjoint 
representation and the fermions and Higgses are in the $27$ 
fundamental representations\cite{e6}.  
A NCQFT model can be constructed following 
the procedures discussed earlier. This model is
very similar to the trinification model with the advantage that no additional
$Z_3$ symmetry is needed. There are however differences and complications. 
In addition to the
gauge bosons in the trinification model, there are also $54$ colored gauge 
bosons. These particles mediate proton decays. Therefore they 
have to be made heavy. To achieve this 
more Higgs representations will have to be introduced which complicate the
theory\cite{e6}. 
The trinification model is simpler in terms of particle contents.

The other comment concerns another approach to construct
trinification model with noncommutative space-time without the use of
Seiberg-Witten mapping adopted in Ref.\cite{11}. In this approach one first
constructs fields in $U(N)$ product groups and then break the symmetry
spontaneously to $SU(N)$ product group. For the trinification model, one can
extend the group to $U(3)_C\times U(3)_L\times U(3)_R$. The gauge field
representation is

\[
27=(9,1,1)+(1,9,1)+(1,1,9). 
\]
In Ref.\cite{11} the symmetry breaking of $U(N)$ to $SU(N)$ is assumed to be
achieved by non-zero VEV's of representations $S_{i}$ which transform as
singlets under the $SU(N)$ but with non-zero charge for $U(1)$ subgroup of $%
U(N)$. Following Ref. \cite{11} one can introduce three $S_{i}$ fields for each
of the factor $U(N)$ group. The VEV's of these fields break the group to the
trinification group discussed earlier producing three heavy gauge bosons.
However it has been shown that models based on such approach violates
unitarity\cite{rh}. This approach may not lead to a realistic model.

This work was supported in part by National Science Council under grants NSC
89-2112-M-002-058 and NSC 89-2112-M-002-065, and in part by the Ministry of
Education Academic Excellence Project 89-N-FA01-1-4-3. I would like to thank
hospitality provided by the Institute of Theoretical Sciences at the
University of Oregon where part of this work was done.

Note Added: Another consistent
noncommutative grand unified model based on $SO(10)$
has been constructed by P. Aschieri et al. (hep-th/0205214), three months
after this work.

\end{document}